\documentclass[12pt]{iopart}

\usepackage{harvard}
\usepackage{graphicx}	
\usepackage{amssymb}	

\newcommand{\aj}{Astron.~J.} 
\newcommand{\apj}{Astrophys.~J.} 
\newcommand{\apss}{Astrophys.~Space~Sci.} 
\newcommand{\aap}{Astron.~Astrophys.} 
\newcommand{\aapr}{Astron.~Astrophys.~Rev.} 
\newcommand{\mnras}{Monthly~Not.~Royal~Astron.~Soc.} 
\newcommand{\prd}{Phys.~Rev.~D} 
\newcommand{\prl}{Phys.~Rev.~Lett.} 

\begin{document}

\title[Cosmological constant from boundary condition]{Cosmological
  constant caused by observer-induced boundary condition}

\author[J. O. Stenflo]{Jan O. Stenflo}

\address{Institute for Particle Physics and Astrophysics, ETH Zurich, CH-8093
Zurich, Switzerland, and\\
Istituto Ricerche Solari Locarno (IRSOL), Via Patocchi, CH-6605 Locarno-Monti, Switzerland}
\ead{stenflo@astro.phys.ethz.ch}

\begin{abstract}
The evolution of the wave function in quantum mechanics is deterministic like
that of classical waves. Only when we bring in observers the fundamentally
different quantum reality emerges. Similarly the introduction of
observers changes the nature of spacetime by causing a split between
past and future, concepts that are not well defined in the observer-free
world. The induced temporal boundary leads to a resonance condition for 
the oscillatory vacuum solutions of the metric in Euclidean time. 
It corresponds to an exponential de Sitter evolution in real
time, which can be represented by a cosmological constant $\Lambda
=2\pi^2/r_u^2$, where $r_u$ is the radius of the 
particle horizon at the epoch when the observer exists. For
the present epoch we get a value of $\Lambda$ that agrees with the
observed value within $2\sigma$ of the observational errors. This
explanation resolves the cosmic coincidence problem. Our epoch in
cosmic history does not herald the onset of an inflationary phase
driven by some dark energy. We show that the observed accelerated
expansion that is deduced from the redshifts is an ``edge effect'' due
to the observer-induced boundary and not representative of the
intrinsic evolution. The new theory satisfies the
BBN (Big Bang nucleosynthesis) and CMB (cosmic microwave background)
observational constraints equally well as the concordance model of standard
cosmology. There is no link between the dark energy and dark matter
problems. Previous conclusions that dark matter is mainly non-baryonic
are not affected. 
\end{abstract}

%
\vspace{2pc}
\maketitle
%


\section{Introduction}\label{sec:intro}
The cosmological constant had to be reintroduced as a free modeling
parameter after the unexpected discovery of
the cosmic acceleration in the end of the 1990s through the use of
supernovae type Ia as standard candles (Riess et al. \citeyear*{stenflo-riessetal1998a},
Perlmutter et al. \citeyear*{stenflo-perlmutteretal1999a}). 
Its small but non-zero positive value has since been
regarded as one of the great mysteries of contemporary physics 
\citeaffixed{stenflo-binetruy2013}{cf.}. It is
some 120 orders of magnitude smaller than one would 
expect for vacuum fluctuations in quantum field theory. Usually it is
interpreted as representing some new kind of physical field,
referred to as ``dark energy'', but its
appearance as a constant would then imply that our present time is
singled out in cosmic history as exceptionally special. The ``Now''
would signal the onset of an 
inflationary phase driven by the dark energy for all eternity,
although in the past this energy was insignificant by many
orders of magnitude. This would contradict the Copernican Principle,
which states that we are not privileged observers. The issue is generally referred
to as the ``cosmic coincidence problem''. 

The possible physical origin of the cosmological constant has
  been addressed from a variety of directions since the discovery of
  the cosmic acceleration, e.g. by
  \citeasnoun{stenflo-velten_etal2014},
  \citename{stenflo-padmanabhan2014}
  \citeyear{stenflo-padmanabhan2014,stenflo-padmanabhan2017}, and
  \citeasnoun{stenflo-lombriser2019}. 
Recently (\citename{stenflo-s2018cc}
\citeyear*{stenflo-s2018cc,stenflo-s2019apss,stenflo-2020intech}) it 
has been shown that the observed value
of the cosmological constant can be predicted
without the use of any free parameters or modifications of Einstein's
equations, if the observed cosmic acceleration is caused by an
observer-induced boundary 
condition instead of by some new kind of physical field. The boundary
constraint affects the redshift pattern that surrounds each observer,
in a way that can 
be described by a cosmological constant. The so 
induced cosmological constant does not herald the onset of a new
inflationary phase, which would make our present epoch special. 

While the idea that the cosmological constant could be related to
boundary conditions is not new
\citeaffixed{stenflo-hayakawa2003,stenflo-banks2018,stenflo-gaztanaga2020}{e.g.},
the earlier treatments are conceptually very different and have not had much
predictive power. What fundamentally sets the present approach apart
is the observer-induced nature of our boundary condition. 
While the analytical expression for the cosmological constant was
  derived in our previous papers on this topic, starting with
  \citeasnoun{stenflo-s2018cc}, the viability of the theory has
  remained undetermined, because the appropriate cosmological
  framework that would allow confrontation with the observational
  constraints has been missing. To address this we need a new
  conceptual interpretation of the meaning and physical implications
  of the observer-induced boundary condition. It is done in the
  present paper. Although the resulting theory is found to be
  conceptually very different from that of standard cosmology, we show
  that the BBN (Big Bang nucleosynthesis) and CMB (cosmic microwave
  background) observational constraints are satisfied equally well as
  by the concordance model (while we also predict the numerical value
  of the cosmological constant, something that concordance cosmology
  is unable to do). 

In Sect.~\ref{sec:bound} we address the origin of the boundary
condition and explain why it is a direct consequence of the presence of 
the observer, who causes a split of the time line between past and
future. To derive the effect that the boundary has on the vacuum modes
of the metric we need to make use of two complementary concepts of
time: real and imaginary time. In Sect.~\ref{sec:facet} we clarify the meaning of
Euclidean or imaginary time and how it is related to thermodynamics
and Hawking's no-boundary proposal for the origin of the
universe. With this background we derive in Sect.~\ref{sec:deriv} the theoretical
expression for the cosmological constant $\Lambda$ that corresponds to
oscillatory vacuum modes of the metric. Comparison with the observed
value for $\Lambda$ shows nearly perfect agreement if $\Lambda$ is determined by the
resonance mode that has a wavelength given by the conformal age
of the universe (in time units) or the radius of the particle horizon
(in spatial units). Such a resonance expresses a periodic boundary
condition across the bounded time line. In Sect.~\ref{sec:period} we
clarify why it is a natural consequence of the observer-induced
boundary constraint, and show 
how it is related to thermodynamics. The resonance condition implies
that the value of the cosmological constant is tied to the conformal
age $\eta_u$ of the universe, such that $\Lambda\sim 1/\eta_u^2$. This
leads to a different cosmological framework with implications for
cosmic history, as explained in Sect.~\ref{sec:history}. For the interpretation of
observational data with this new framework we need to distinguish
between the ``intrinsic'' evolution of the universe, and the ``edge
effects'' that the observer sees in the observed redshift-distance
relation. These concepts are explained in Sect.~\ref{sec:intrin} together 
with plots that illustrate how the redshift pattern in the vicinity of
each observer is modified by the observer-induced boundary. In a local
region or ``bubble'' around the observer, who defines the ``edge'', one gets 
enhanced and accelerated expansion with properties that are described
by the induced cosmological constant. In Sect.~\ref{sec:baryon} we explain why the
new cosmological framework satisfies all the BBN and CMB 
observational constraints. We show that there is no link between the
dark energy and dark matter problems. The long-standing conclusion that
most of dark matter must be non-baryonic is not affected. Section \ref{sec:concl}
presents the concluding remarks.

\section{Boundary condition from observer
  participation}\label{sec:bound} 
To understand the origin of the boundary condition it is helpful to
draw a parallel with the measurement problem of QM (quantum mechanics). The
result of a measurement depends on the experimental or observational
framework. While the underlying equations are not changed, the set-up
constrains the solutions of the equations. Heisenberg's uncertainty
principle is a consequence of this. For instance, if the set-up
constrains the accessible time interval, 
energy fluctuations will be induced. There is complementarity between
spacetime and momentum-energy, between waves and particles, between
the real domain and its Fourier counterpart. 

In Einsteinian GR (general relativity) the space and time dimensions
extend indefinitely. Time has no boundaries (except singularities,
like the Big Bang in the cosmological case). This describes
an objective, classical world that ignores the participatory role
of observers. It makes GR incompatible with QM, where observer
participation profoundly changes the nature of reality (collapse of
the wave function, probabilistic causality, etc.). 

The introduction of an observer brings a profound change
to classical GR by implying a split between
past and future, concepts that have no well-defined meaning in a world
without observers. The presence of the observer unavoidably changes
the observational framework. The time line gets 
truncated, because the future is not accessible, even in
principle. Instead of dealing with an infinite time dimension, time is
now bounded between the Big Bang singularity and the observer-defined
new edge, the Now. In QM a finite time interval affects the 
vacuum energy. The bounded cosmological time line has analogous
consequences (although they are different, as clarified in Sect.~\ref{sec:period}). 

In the next sections we will show how the observer-induced temporal
boundary affects the vacuum wave modes of the metric. Thus only 
modes with periodic boundary conditions in the Euclidean
(imaginary time) representation are allowed. Oscillatory modes in
imaginary time correspond to de Sitter type exponential evolutions in
real time. The resonance condition, which follows from the boundedness
of the time line, implies the existence of a $\Lambda$ term with a magnitude
that is found to agree with the observed value within  about 2\,\%,
without the use of any free parameters. 

There has been considerable controversy about the use and physical
meaning of imaginary time, after \citeasnoun{stenflo-hartlehawking83} introduced
it to eliminate the initial singularity. Hawking has repeatedly
referred to this idea as the ``no-boundary proposal'' for the origin
of the universe (cf.~\citename{stenflo-hawking_bcuniv82} 
\citeyear*{stenflo-hawking_bcuniv82,stenflo-hawking_qmstate84}). 
He nevertheless remained unclear about the physical reality 
of imaginary time vs. its instrumentalist use as a mathematical tool
without real physical meaning
\citeaffixed{stenflo-deltete_guy96}{cf.}. Our application of imaginary
time to 
successfully predict the observed value of the cosmological constant
leads us to a somewhat different viewpoint, which is
conceptually similar to the wave-particle complementarity in QM.

\section{The multifaceted nature of time}\label{sec:facet}
The metric of a homogeneous, isotropic, and flat universe can be
expressed in terms of three time concepts: proper time $t$, conformal
time $\eta$, and Euclidean conformal time $\tau$: 
\begin{eqnarray}\label{eq:etataumetric}
{\rm d}s^2 &=& -c^2 {\rm d}t^2 +a(t)^2 \,({\rm d}r^2+ \,r^2\,{\rm
  d}\Omega)\,,\nonumber\\ 
{\rm d}s^2 &=& \,a(\eta)^2 \,(\,-c^2 {\rm d}\eta^2 +\,{\rm d}r^2+ \,r^2\,{\rm
  d}\Omega)\,,\\ 
{\rm d}s^2 &=& \,a(\tau)^2 \,(\, {\rm d}\tau^2 +\,{\rm d}r^2+ \,r^2\,{\rm d}\Omega)\,.\nonumber
\end{eqnarray}

The Lorentzian representation of the metric with signature ($-++\,+$)
is generally considered as the ``correct'' representation,
because it allows a covariant description with a geometric
interpretation, according to which the Lorentz transformations have
representations as rotations of 4D spacetime. The 4D ``marriage'' of
time with space that was introduced by Minkowski has been so powerful that
it has led to the view that time is just another dimension like
space (albeit with opposite sign in the metric signature). However,
this view of time is rather restricted, because it overlooks the circumstance that
time is profoundly different from space in several other ways: it has
an arrow, future is not observable, time travel is impossible. In
addition, 
time has a deep connection with thermodynamics, entropy, and the 2nd
law. Since these fundamental aspects are not contained in the Lorentzian
geometric description, it is incomplete and needs to be complemented by
other representations. Euclidean or imaginary time is such a 
representation. It is also incomplete, because it does not allow a
geometric interpretation of the Lorentzian coordinate transformations,
but it is nevertheless physically relevant, because it provides a
representation of other profound aspects. 

The Euclidean representation reveals a deep connection between
statistical mechanics and QFT (quantum field theory) and has been
widely used, in the form of Euclidean field theory, to deal with
critical phenomena and phase transitions in condensed matter physics
\citeaffixed{stenflo-zee2010,stenflo-euclft2011}{cf.}. It also exposes a deep link between 
general relativity and thermodynamics, in a way that opens up a direct
route to derive the temperature of black holes
\citeaffixed{stenflo-gibbonsperryprl78,stenflo-zee2010}{cf.}. Hawking 
saw its potential for a theory of quantum gravity, a theory without
singularities. This is the idea and motivation behind his
``no-boundary proposal'' (cf.~\citename{stenflo-hawking_bcuniv82} 
\citeyear*{stenflo-hawking_bcuniv82,stenflo-hawking_qmstate84}).

We know that GR is incomplete, because it contains singularities (like
Big Bang or the center of black holes), where the theory breaks down
\cite{stenflo-penrose65,stenflo-hawkingpenrose70}. It has been shown 
\cite{stenflo-hartlehawking83,stenflo-hawking_qmstate84} that 
this break-down may be avoided with the help of an imaginary-time
representation. Hawking's viewpoint has been that time becomes
imaginary in the trans-Planckian region as we approach the initial
singularity. This has the consequence that the
physical singularity gets transformed into a benign coordinate
singularity (like that of the spherical coordinates at the north or south
poles). Around the Planck era there would be a transition of time
from imaginary to real, after which the cosmic evolution would proceed
in the standard manner. 

A major problem with Hawking's viewpoint is 
that the mechanism, which is responsible for the transition from imaginary to real time,
the so-called ``join problem'' \citeaffixed{stenflo-deltete_guy96}{cf.}, has never
been specified. As will be argued in the following
sections, the join problem never arises  in our different
interpretation of imaginary time: both real and imaginary time are
physically valid but 
complementary aspects of the same underlying reality. The situation is
analogous to the complementarity between waves and particles in
QM. Although seemingly incompatible, they are both valid complementary
aspects of physical reality. In terms of this analogy, Euclidean time
represents the wave aspects (because it allows a QFT
representation with oscillating phase factors), while real time
corresponds to the particle aspects (with the energy-momenta that
drive the cosmic expansion). We may also draw an analogy with optics,
where the complementary aspects of absorption and dispersion (or
amplitude and phase) are unified with the help of a complex refractive index.

\section{Derivation of the expression for $\Lambda$}\label{sec:deriv}
The conformal metric preserves all the angles and relations between
spatial and temporal coordinates, including the light-cone structure
of Minkowski spacetime. We need the conformal framework for the
description of the global Fourier modes of the metric in a way that is
unaffected by the differential distortions by the scale factor
$a$. 

Allowing for a cosmological constant $\Lambda$ (of so far undefined
magnitude), the vacuum 
modes of the metric are governed by  
\begin{equation}\label{eq:einst}
R_{\mu\nu} -\Lambda\, g_{\mu\nu}=0\,. 
\end{equation}
In the case of a homogeneous universe without spatial gradients, the
weak-field approximation in the harmonic gauge
\citeaffixed{stenflo-weinberg72}{cf.}  gives us 
\begin{equation}\label{eq:weakfield}
\frac{1}{2\,c^2}\,\frac{\partial^2
g_{\mu\nu}}{\partial \eta^2} -\Lambda\, g_{\mu\nu} =0\,.
\end{equation}
The choice of a particular gauge does not restrict the validity
  of the equation, because such a choice merely represents a mathematical
  procedure to deal with redundant degrees of freedom to simplify the
  equations. In general relativity like
  in all of classical physics and abelian quantum physics, the choice
  of gauge has no physical significance. 

Equation (\ref{eq:weakfield}) 
does not have any oscillatory solutions when $\Lambda$
is positive (which it is observed to be), because the two terms in the
equation have opposite signs. Wave solutions only exist if we
switch to imaginary (Euclidean) time $\tau$, because the first term then
changes sign to give us the equation for a harmonic oscillator.  

It is convenient to convert Eq.~(\ref{eq:weakfield}) to the standard form of a
harmonic oscillator:   
\begin{equation}\label{eq:harosc}
-\,\frac{\partial^2 g_{\mu\nu}}{\partial \eta^2} +\,2c^2\,\Lambda\, g_{\mu\nu} =c^2\,\frac{\partial^2 g_{\mu\nu}}{\partial \tau^2} +\omega^2_\Lambda\, g_{\mu\nu} =0\,,
\end{equation}
where we have introduced the frequency $\omega_\Lambda$ of the
harmonic oscillator that satisfies the equation. Identification
gives 
\begin{equation}\label{eq:lamomtlam}
\Lambda= \,\frac{\omega^2_\Lambda}{2c^2}\,.
\end{equation}
Since 
\begin{equation}\label{eq:omlam}
\omega_\Lambda=\frac{2\pi}{\eta_\Lambda}\,,
\end{equation}
where $\eta_\Lambda$ is the period of the oscillations, we get 
\begin{equation}\label{eq:lametalam}
\Lambda= 2\left(\frac{\pi}{c\,\eta_\Lambda}\right)^{\!\!2}\!\!.
\end{equation}

Let us use subscript $u$ to mark that the
respective quantity refers to a certain epoch (when the age of the
universe is $t_u$ and the scale factor is $a_u$).  If we for
$\eta_\Lambda$ insert the value $\eta_u$ of the present conformal age that is 
derived from the concordance model, then 
Eq.~(\ref{eq:lametalam}) gives us a value of
$\Lambda$ that agrees with the observed value within the observational
errors. 
Note that the agreement is not just within an order of magnitude but within 2\,\%\ of the observations. The implied equality 
$\eta_\Lambda \approx\eta_u$ is too precise 
to be dismissed as a fortuitous coincidence. It convincingly suggests
that Eq.~(\ref{eq:lametalam}) must represent a resonance
condition for the vacuum mode of $g_{\mu\nu}$ across the bounded
$\tau$ interval (between the Big Bang and the Now).  In the next section
we will address the origin of this resonance.

Instead of temporal units we can express the resonance in terms of
spatial units with the help of the radius $r_u$ of the so-called
particle horizon. It is related to the conformal age through
$r_u=c\,\eta_u$. This gives us the simple formula 
\begin{equation}\label{eq:lamru}
\Lambda_u= 2\left(\frac{\pi}{r_u}\right)^{\!\!2}
\end{equation}
for the cosmological constant, which applies to any epoch of cosmic
history.

\section{Periodic boundary condition and
  thermodynamics}\label{sec:period} 
Because the exponential (de Sitter) evolution in real time corresponds to an
oscillating phase factor in Euclidean (imaginary) time, we can
represent Euclidean time in terms of an angular coordinate. Due to
the presence of the 
observer, the physically meaningful Euclidean time string $\tau_u$ is
bounded. Its length is 
the age of the universe in conformal Euclidean time units. We can
express it in terms of angular units such that
its length is $2\pi$. This scaling corresponds to the application of periodic boundary
conditions. The resonance can be expressed geometrically by wrapping
$\tau_u$ around the unit circle  exactly once. Such a 
circular ``Euclidean world'' is bounded while having no boundaries. 
In real, spatial units, the period or wavelength of the resonance
equals $r_u$. Equation (\ref{eq:lamru}) is therefore a 
consequence of applying periodic boundary conditions to the bounded
$\tau_u$ interval, or, in terms of the geometric picture, wrapping it once
around the circle. 

According to Euclidean field theory with its powerful applications in
solid-state physics \citeaffixed{stenflo-euclft2011}{cf.}, 
the phase factor across the finite
interval with periodic boundary conditions in Euclidean time
becomes the Boltzmann factor in real time. Thus 
\begin{equation}\label{eq:euclidtemp}
e^{\,i\,\omega_u\,\tau_u/c} = e^{\,-\,\omega_u\,\eta_u} \equiv e^{\,-\hbar\, \omega_u/(k_B T_u)}\,,
\end{equation}
which implies that the bounded time string induces a temperature 
\begin{equation}\label{eq:tempu}
T_u =\frac{\hbar}{k_B \,\eta_u}\,.
\end{equation}
The equipartition theorem tells us that each degree of freedom of the
system has energy $\Delta E = {\textstyle\frac{1}{2}}\,k_B
T_u$. Denoting the restricted time interval by $\Delta t \equiv
\eta_u$, we see that Eq.~(\ref{eq:tempu}) is equivalent to the
Heisenberg relation 
\begin{equation}\label{eq:heisen}
\Delta E\,\Delta t = {\textstyle\frac{1}{2}}\,\hbar\,.
\end{equation}
It expresses how energy fluctuations are induced in the system when
the time interval that is accessible to the observer is
restricted. 

Due to the large magnitude of $\eta_u$, the present value of $T_u$ is
about $10^{-29}\,$K, which is entirely insignificant as compared with
the ambient CMB temperature. It is much too small to play a role in
driving the observed cosmic acceleration and should not be confused
with the apparent energy density that is embodied by the cosmological
constant $\Lambda$. Although both $T_u$ and
$\Lambda$ are induced by a finite time interval (the age of the
universe), they are fundamentally different. In particular the
expression for $\Lambda$ does not contain 
Planck's constant $\hbar$. It is therefore not directly related to
quantum physics. 

The mode energy $\hbar\,\omega_u$, which
indicates how much the metric is disturbed, is
about $10^{-61}$ in Planck units (because $\eta_u$ is about $10^{61}$
in units of the Planck time). This justifies the use of the weak-field
approximation, because we are dealing with fluctuations of very
small energy. The magnitude of $\Lambda$, however, does not
depend on the fluctuation amplitude (which scales with $\hbar$) but
depends exclusively on the value of the resonance frequency $\omega_u
=2\pi/\eta_u$.

\section{Implications for cosmic history}\label{sec:history}
In cosmology it is convenient to express the matter and radiation
energy densities of the stress-energy tensor $T_{\mu\nu}$ on the
right-hand 
side of the Einstein equation in terms of the dimensionless parameters
$\Omega_M$ and $\Omega_R$, which represent the respective energy
densities in units of the critical energy density that separates open
and closed model universes. We are free to move the $\Lambda_u$ term
to the right-hand side and interpret it as a fractional mass-energy
density $\Omega_\Lambda$ given by 
\begin{equation}\label{eq:omlamau}
\Omega_\Lambda(a_u) = \frac{c^2}{3H_u^2}\,\Lambda (a_u)\,.
\end{equation}
$a_u$ is the local (at redshift $z=0$) scale factor $a(t_u)$ for the epoch of
the given observer. Present-day 
observers are at epoch $t_u=t_0$, when the Hubble constant $H_u=H_0$,
while the scale factor $a_u=a_0=1$ is normalized to unity. 

In standard cosmology $\Lambda$ is a true constant that is independent
of epoch and $a_u$. This is not the case when $\Lambda$ is induced by our periodic
boundary condition, because it tracks the age or
size of the observable universe as expressed by
Eq.~(\ref{eq:lamru}). In spite of this tracking property it is
important to understand that both $\Lambda(a_u)$ and
$\Omega_\Lambda(a_u)$ do not depend on redshift $z$ from the perspective
of the observer at epoch $a_u$. This may at first seem strange but becomes
clear if we realize that $\Lambda(a_u)$ and $\Omega_\Lambda(a_u)$ do
not represent physical fields but are emergent quantities from a
boundary constraint. Let us clarify. 

The values of $\Lambda(a_u)$ and $\Omega_\Lambda(a_u)$ are tied to the 
global resonance frequency $\omega_u$, which applies to the totality
of the observable universe at the epoch that is defined by $a_u$. The
musical tones that emanate from a violin string do not depend on
position along the string, they represent a global property of the
oscillating system. Quantum numbers do not vary with position within
the atom but represent resonances of the system. Similarly,
$\omega_u$, $\Lambda_u$, and $\Omega_{\Lambda_u}$ do not vary with
spacetime position across the observable universe that is defined by
epoch $a_u$. Therefore, although they vary with $a_u$ (which is a
quantity that exclusively refers to redshift zero), they do not vary
with redshift. When we change $a_u$, we shift the zero point of the
whole redshift scale (because the observer in an expanding universe is
by definition always located at zero redshift). 

If $\Lambda$ were due to some physical field, then it would need
  to be independent of epoch and $a_u$ (as it is in standard cosmology),
  because this is demanded by energy conservation together with the
  Bianchi identities. Since $\Lambda_u$ is independent of redshift for
  any given epoch, the Bianchi identities are indeed satisfied across
  the entire 4D spacetime that is observable at that epoch. The
  boundary conditions (and consequently the cosmological constant)
  change when we move to a different observer epoch, but this change
  does not violate the Bianchi identities, because the boundary
  conditions are not governed by some continuity or energy
  conservation equation. Instead they are governed by the global resonance
  condition that depends on the conformal age of the universe. 

The solution of the Einstein equation for an isotropic, flat, and
homogeneous universe can be written as 
\begin{equation}\label{eq:hhu}
H=H_u\,E_u(y)\,,
\end{equation}
where 
\begin{equation}\label{eq:ydef}
y\equiv a/a_u=1/(1+z)
\end{equation}
and 
\begin{equation}\label{eq:ezy}
E_u(y)=[\,\Omega_M(a_u)\,y^{-3} +\,\Omega_R(a_u)\,y^{-4} +\,\Omega_\Lambda(a_u)\,]^{1/2}\,.
\end{equation}

Let us for convenience define the dimensionless conformal age $x_u$ of the
universe as the conformal age $\eta_u$ in units of the Hubble time
$1/H_u$. Then 
\begin{equation}\label{eq:xuinteg}
x_u \equiv\eta_u\,H_u=H_u\!\int_0^{t_u} \!\frac{{\rm d}t}{ (a/a_u)}=\int_0^1\!
\frac{{\rm d}y}{y^2\,E_u(y)}\,.
\end{equation}
With Eqs.~(\ref{eq:lamru}) and (\ref{eq:omlamau}) it gives us a
compact expression for $\Omega_\Lambda$: 
\begin{equation}\label{eq:omlamtheo}
\Omega_\Lambda(a_u)\,=\,\frac{2}{3}\left(\frac{\pi}{x_u}\right)^{\!\!2}. 
\end{equation}

This seemingly innocent equation cannot be solved directly, because
$x_u$ on the right-hand side depends on $\Omega_\Lambda(a_u)$ on the
left-hand side. The value of $x_u$ is calculated from the $E_u(y)$ function,
which in turn needs $\Omega_\Lambda(a_u)$ to be defined. Equation
(\ref{eq:omlamtheo}) can however be solved by straightforward
iteration. The procedure is described in detail in \citeasnoun{stenflo-2020intech}. The
iterations converge quickly to a unique solution. 

In particular, the iterative solution for the present epoch 
verifies that we get nearly perfect agreement with the observed
value for $\Omega_\Lambda$. Our 
theory gives us (without the use of any free fitting parameters) $\Omega_\Lambda
=67.2\,$\%, which is within $2\sigma$ from the value $68.5\pm
0.7\,$\%\ that has been derived from data with
the Planck satellite (Planck Collaboration et al. \citeyear*{stenflo-planck2018arX}).

While $\Lambda$ is a constant that is independent of $a_u$ in standard
cosmology, the value of the dimensionless parameter $\Omega_\Lambda$
scales as $1/H_u^2$ and therefore varies steeply with $a_u$. It
was insignificant in the past but will dominate in the future. We
happen to live at an epoch when $\Omega_\Lambda$ is comparable in
magnitude to $\Omega_M$ (which constitutes the cosmic coincidence problem). 

In contrast, $\Omega_\Lambda$ as derived from our cosmic boundary
constraint has a value that does not change with $a_u$ unless the
equation of state of the matter-radiation content of the universe
changes. For a universe with only matter and no radiation we would
have $\Omega_\Lambda=66.3\,$\%\ (which is somewhat different from the
$67.2\,$\%\ that we obtain for the present epoch, because the
contribution to the $x_u$ integral from $\Omega_R$ is not
negligible). For a universe with no matter but only radiation (which
is a good approximation for the early universe) we get
$\Omega_\Lambda=93.1\,$\%\ instead.

\section{Intrinsic evolution and edge effects in the redshift
  pattern}\label{sec:intrin} 
In quantum mechanics the evolution of the wave function that is
described by the Schr\"odinger equation represents an evolution that is
undisturbed by the presence of observers. It describes a world that is
deterministic with well-defined causality and no uncertainty. Let us
for convenience refer to it as the ``intrinsic'' evolution. It is
unobservable. Once an observer is introduced, the description changes
profoundly. 

In analogy, the ``intrinsic'' evolution of the scale factor $a(t)$ in
cosmology is governed by the Einstein equation for a world without
observers. The insertion of an observer implies the introduction of a
temporal boundary, an edge that generates a repulsive force as
expressed by the induced $\Lambda$ term. 
Note that the observer can be inserted at any time, there is
  nothing special about the present epoch. The cosmological constant
  is an observer-induced effect for any epoch in cosmic history. 

The repulsive force has two main effects: (1)
It enhances the expansion rate in the vicinity of the edge (i.e., for
small redshifts). (2) It accelerates the expansion 
in a local region or ``bubble'' that surrounds the observer. 
These are ``edge effects'' caused by the presence of a boundary. They
are large in the vicinity of the boundary but vanish at large
distances from it. They manifest themselves as a change in the local  redshift
pattern, a change that becomes
insignificant for large redshifts. Let us now explore them 
in quantitative detail. 

Retaining the assumption of zero spatial curvature, the intrinsic 
cosmological evolution without the presence of observers is a
Friedmann universe, obtained from the concordance model by removal of 
the cosmological constant. The matter $M$, radiation $R$, and $\Lambda$
terms each contribute to the squared expansion rate $H^2$ with amounts
$H_0^2 \,\Omega_{M,R,\Lambda}$. If we remove the $\Lambda$
contribution, the expansion rate will be reduced. Let us use subscript
$F$ to distinguish the intrinsic, underlying Friedmann model from the
model without subscripts that describes the observed universe. Then
$H_F < H$ in the vicinity of the observer. If $H_{F_0}$ denotes the
value of $H_F$ for $a=a_0=1$, then 
\begin{equation}\label{eq:hf02}
H_{F_0}^2=H_0^2\, (\Omega_M +\Omega_R)=H_0^2\, (1-\Omega_\Lambda)\,,
\end{equation}
where the values of $H_0$ and $\Omega_{M,R,\Lambda}$ are
determined by the observational constraints (Planck Collaboration et
al. \citeyear*{stenflo-planck2018arX}). 
As an alternative one can use the
nearly identical value of 
$\Omega_\Lambda$ that is obtained from Eq.~(\ref{eq:omlamtheo}) of our
theory, which automatically also gives us the value of
$\Omega_M\approx 1-\Omega_\Lambda$ that follows because $\Omega_R\ll
1$. Accordingly 
\begin{equation}\label{eq:hf0}
H_{F_0}=H_0\,\sqrt{1-\Omega_\Lambda}\,,
\end{equation}
which uniquely defines the ``intrinsic'', underlying (and
unobservable) Friedmann model. It is (through its definition) fully
consistent with the 
presently observed values for $H_0$ and $\Omega_\Lambda$. 

\begin{figure}
 \includegraphics[width=\columnwidth]{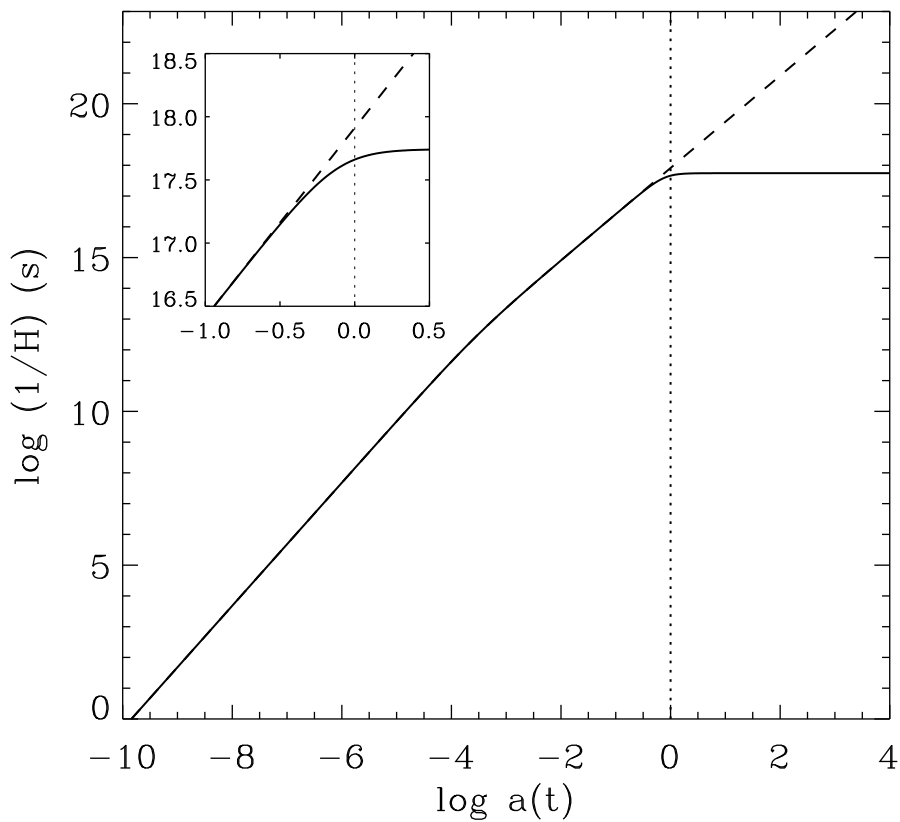}
\caption{Evolution in a log-log representation of the inverse
  expansion rate $1/H$ 
  vs. scale factor $a$. The solid curve represents the
  concordance model of standard cosmology, the dashed curve the
  governing but unobservable intrinsic Friedmann model. The dotted
  vertical line marks the present epoch. According to
  the concordance model the Now is singled out as something very
  special, with an abrupt transition to an eternal inflationary phase. This
  transition is shown in greater detail in the inset diagram. There is
  nothing special about our present time according to the intrinsic
  model. 
}\label{fig:hvsa}
\end{figure}

Figure \ref{fig:hvsa} shows the evolution of the inverse expansion
rate of the intrinsic Friedmann model (dashed), compared with the
present concordance model of standard cosmology (solid). The most
striking feature of this plot is the abrupt onset of an inflationary
phase for the concordance model, an exponential expansion with a rate
that does not depend on the scale factor. We happen to live at a time
in cosmic history when this sudden transition takes place. It has been 
considered as a profound mystery why our epoch is singled
out for such a dramatic cosmic event. 
This mystery does not exist in our cosmology, because the predicted
inflationary phase is an artefact of incorrectly interpreting the edge effect as
a physical field (dark energy). 

With our explanation of the cosmological constant as a
boundary-induced effect we instead find that our epoch is not different from any
other epoch, we are not privileged observers. The cause of the
abrupt transition of the solid curve is the
insertion of an observer (us) at the present epoch. It 
creates a boundary between past and future, concepts that do not have
any precise meaning 
in the intrinsic, observer-free model. The strange apparent onset of
an inflationary phase does not represent what will
happen in the future, because the future portion of the diagram 
lies beyond the temporal boundary and therefore has no physical meaning. 

The difference between the concordance and intrinsic models 
vanishes the further we look back in time. As seen in Fig.~\ref{fig:hvsa}, the
two models are practically indistinguishable for $\log a \lesssim -1$. 
Near the epochs of decoupling (CMB formation, around $\log a\approx -3$)
and element synthesis (BBN processes, around $\log a\approx -9$) there
is no significant difference between the expansion rates of
the two models. The cosmology that follows from our boundary-induced
explanation of the cosmological constant therefore satisfies all BBN
and CMB observational constraints in the same way and with equal
precision as concordance cosmology. 

\begin{figure}
 \includegraphics[width=\columnwidth]{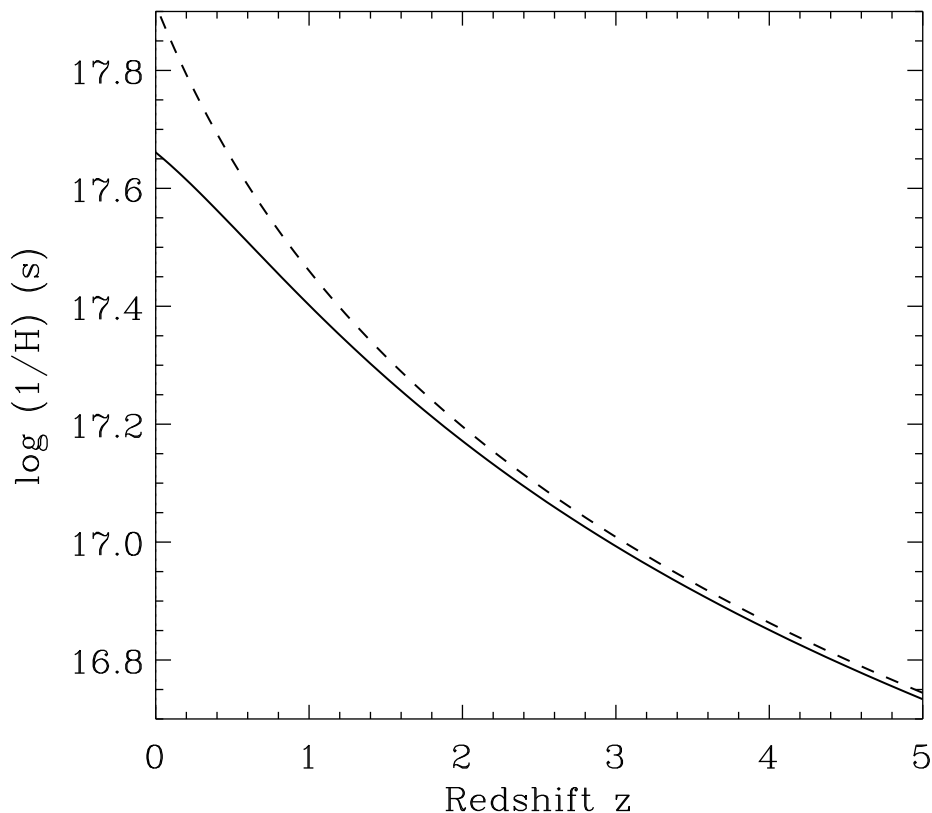}
\caption{Log of the inverse expansion rate $1/H$ vs. redshift $z$ for the
  concordance model (solid) and the intrinsic Friedmann model
  (dashed). The information content is the same as that of Fig.~\ref{fig:hvsa}
  for the observable portion of the universe (where $a<1$). The
  enhancement of the expansion rate (which lowers the solid curve relative to
  the dashed one) is an edge effect, because it vanishes asymptotically for
  redshifts $\gtrsim 3$. 
}\label{fig:hvsz}
\end{figure}

Figure \ref{fig:hvsa}  is not a suitable representation of what the
observer sees. The future region of the diagram (where $\log a >0$) is not
observable. The scale factor $a$ in the past region is only
indirectly observable in the form of redshift $z$, because
$1+z=1/a$ for $a\le 1$. In Fig.~\ref{fig:hvsz} we have therefore
replotted the observable portion of Fig.~\ref{fig:hvsa} vs. redshift
$z$. It better illustrates that we are dealing with an edge effect. In
redshift space the edge is always located at $z=0$, regardless of the
choice of epoch $t_u$ for the observer. All the observed cosmic
acceleration relates 
to the edge (the observer at $z=0$), not to the intrinsic time in cosmic
history, in contrast to the traditional interpretation of standard
cosmology.  Regardless of when in
cosmic history the observations are made, our hypothetical observer 
will find a redshift-distance relation that looks as if an
inflationary phase suddenly begins at that very epoch. Although 
we choose an arbitrary epoch with
our ``Gedankenexperiment'' , the use of standard cosmology for the
interpretation will always make the observer's epoch seem
extraordinarily special. 

As seen from Fig.~\ref{fig:hvsz}, the edge effect manifests itself as
an enhanced expansion rate in the vicinity (for $z\lesssim 3$) of 
the observer (whose existence defines the location of the
boundary). It is an edge effect, because the enhancement
vanishes asymptotically for
redshift $\gtrsim 3$. Note that the enhanced region does not have
any location in the underlying, observer-free intrinsic
spacetime. While the temporal location (the epoch) of the observer
defines the boundary, the location in space is arbitrary. Regardless
of spatial location, the observer is always surrounded by an
enhanced expansion pattern in redshift space. 

\begin{figure}
 \includegraphics[width=\columnwidth]{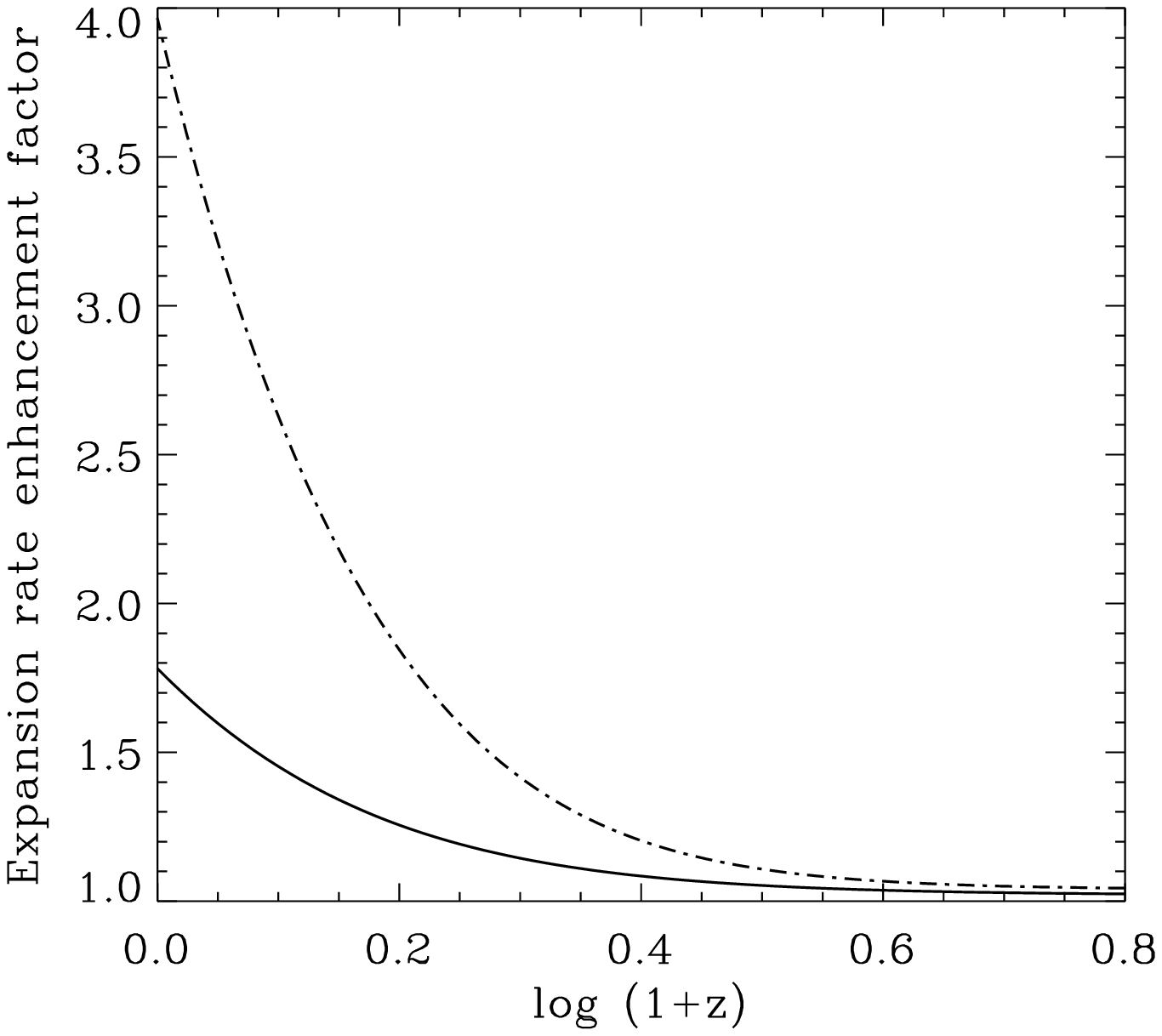}
\caption{Relative expansion rate enhancement, defined as the ratio
  between the observed and intrinsic Hubble constants, plotted against 
  $\log (1+z)$ for two epochs: The Now (solid curve), and the epoch
  when the photon temperature was $10^9\,$K (dash-dotted curve). The
  enhancement was larger in the radiation-dominated era, because the
  boundary-induced value of $\Omega_\Lambda$ was larger. 
}\label{fig:henhanz}
\end{figure}

The magnitude of the edge enhancement effect is exclusively a function
of the value of the boundary-induced dimensionless parameter
$\Omega_\Lambda(a_u)$, which is given by Eq.~(\ref{eq:omlamau}). It only
depends on the equation of state of the matter-radiation content of
the universe. In the early, radiation-dominated universe
$\Omega_\Lambda(a_u)\approx 93.1\,$\%, which is substantially larger
than the present value of $67.2\,$\%. A larger value
of $\Omega_\Lambda(a_u)$ induces a larger edge effect. Figure
\ref{fig:henhanz} illustrates this by giving the relative enhancement,
i.e., the ratio between the observed and intrinsic Hubble constants
$H$, as a function of redshift for two different epochs, the present
(solid), and the BBN epoch when the photon temperature was 1\,GK (dash-dotted). 

The edge enhancement is a local phenomenon that does not
affect the BBN or CMB physics, because the dynamical time scale that
governs the BBN and CMB non-equilibrium processes is determined by the
intrinsic (and unenhanced) evolution. This will be clarified 
in the next section. 

\begin{figure}
 \includegraphics[width=\columnwidth]{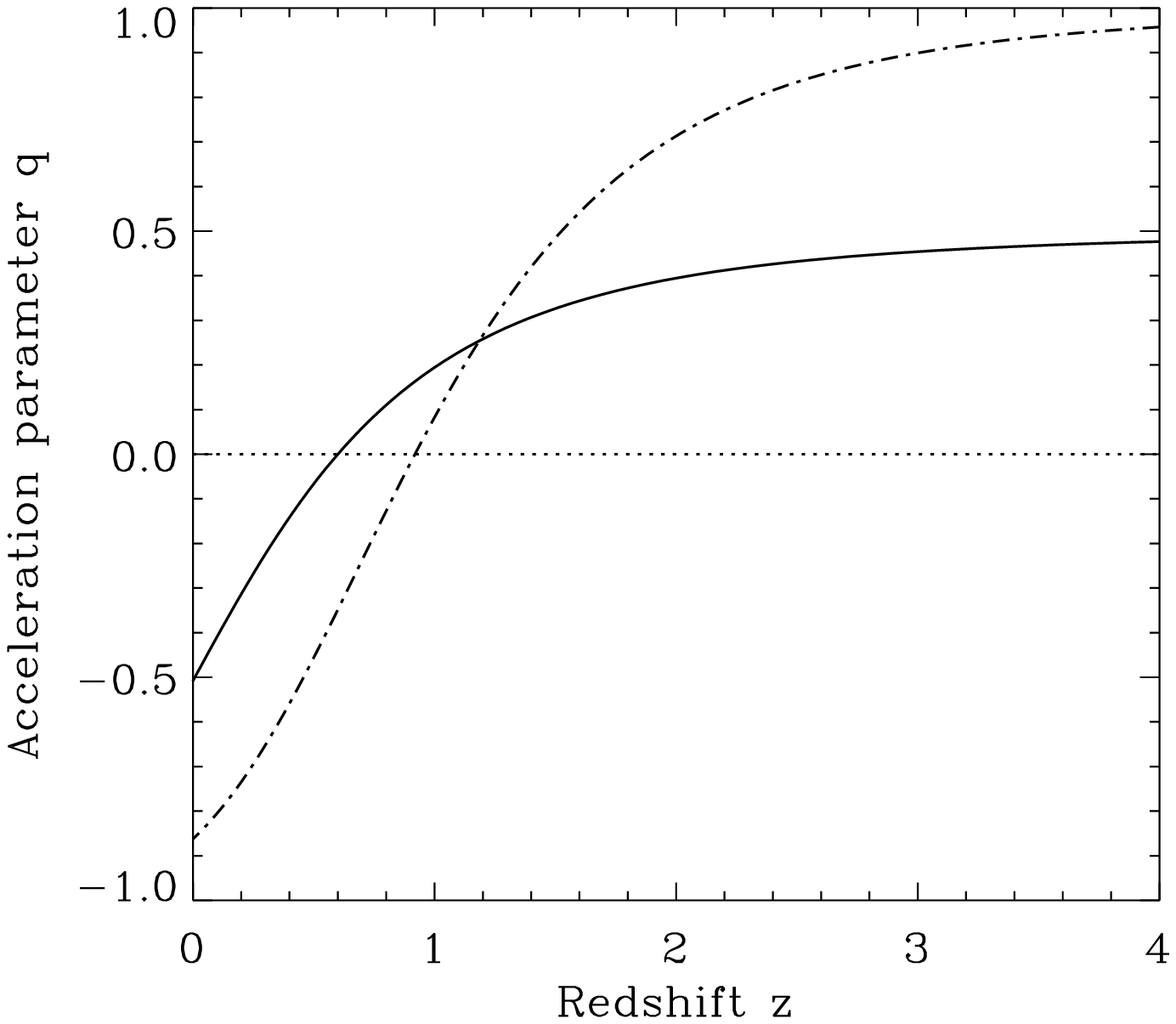}
\caption{Acceleration parameter $q$ that is the result of the
  observer-induced edge effect, plotted vs. redshift $z$ for the same
  epochs as in Fig.~\ref{fig:henhanz}: the Now (solid curve), and the epoch
  when the photon temperature was $10^9\,$K (dash-dotted
  curve). Negative values of $q$ mean acceleration, positive
  deceleration. The diagram shows that each observer is surrounded by
  an accelerating ``bubble'', which extends out to a redshift of about
  0.6 and 1, respectively, for the present and for a radiation-dominated
  universe. 
}\label{fig:qaccz}
\end{figure}

The edge effect not only manifests itself by enhancing the expansion
rate, but also by accelerating it. 
An observer existing at an epoch referred to with index $u$, when
  the scale factor $a=a_u$, would find a cosmic acceleration parameter
$q=q_u(z)$ with a redshift distribution given by 
\begin{equation}\label{eq:qbubble}
q_u(z)=\frac{0.5\,\Omega_M(a_u)(1+z)^3 +\Omega_R(a_u)(1+z)^4-\Omega_\Lambda(a_u)}{\Omega_M(a_u)(1+z)^3 +\Omega_R(a_u)(1+z)^4+\Omega_\Lambda(a_u)}\,.
\end{equation}
Figure \ref{fig:qaccz} shows
the properties of this function for the same two epochs that were illustrated in
Fig.~\ref{fig:henhanz}: the Now, and the epoch when the photon
temperature was 1\,GK. It illustrates that the observer is always
surrounded by an accelerating ``bubble'', not only as discovered from
supernovae observations for the present epoch (represented by the
solid curve in the figure), but even more so in the early universe,
because the observer-induced value of $\Omega_\Lambda(a_u)$ was larger
then. As we have acceleration where $q$ is negative, the radius (in
redshift space) of the accelerating ``bubble'' is given by the $z$ for which
$q$ has a zero crossing. This radius 
is larger (extending to $z\approx 1$) for a
radiation-dominated universe than for the present universe (for
which it extends to $z\approx 0.6$).  
Beyond the bubble the acceleration parameter $q$ quickly approches
the value it would have in the absence of observers, i.e., 
unity for a radiation-dominated, $\textstyle\frac{1}{2}$ for a 
matter-dominated universe.  

Although every observer is forever embedded in an accelerating
region, this is a local (edge) phenomenon, which by itself does not
properly solve the large-scale isotropy and homogeneity problem that
has been the motivation for postulating the occurrence of a
violent GUT-era inflation in the early universe. While there are 
other profound reasons to question the inflation hypothesis
(cf.~\citename{stenflo-penrose04rtr} \citeyear*{stenflo-penrose04rtr,stenflo-penrose16}), 
this topic is outside the scope of the present work.

\section{Confirmation of the need for non-baryonic dark
  matter}\label{sec:baryon} 
In the preceding section we interpreted the boundary-induced
enhancement and acceleration of the expansion as a local phenomenon
that determines how the redshift pattern of the receding
galaxies appears to the observer. It is an edge effect that does not
alter the dynamical time
scale that governs the non-equilibrium physical processes. As will be
clarified below, the relevant physical time scale is the 
expansion time scale of the intrinsic model, which is free from
observer effects. Because the expansion 
rate of the intrinsic model does not differ significantly from the
concordance model during the BBN and CMB eras, both models will
satisfy all observational constraints equally well. In particular, the
conclusion that the fractional baryonic matter content as expressed by the
parameter $\Omega_B$ is only about 5\,\%\ is unaffected by our
boundary-induced explanation of the cosmological constant. 

While it may appear natural, as we have done, to pick the time scale of the intrinsic
model to represent 
the dynamical time scale, this choice needs to be questioned and
properly justified, because it is neither obvious nor unique. It could
for instance be argued that the edge 
effects should be included, that the dynamical time scale should be
determined by the expansion rate that the observer 
experiences directly. Such a modified time scale would differ significantly
from the time scale of the concordance model, which would have major consequences
for the confrontation with the observational data: Would such an
interpretation be compatible 
with all observational constraints\,? If this would turn out to be the
case, would the parameter fits give a
different value for $\Omega_B$\,? In particular, is it conceivable in
principle to come up with a scenario, in which all dark matter would be
baryonic\,? 

The short answer to these questions is that a significantly changed cosmic
expansion history would conflict with the observations. 
The observational constraints are only compatible with one of the
options for the 
dynamical time scale, namely the time scale of the intrinsic,
observer-free model. Alternative interpretations can be 
ruled out. We now know that the alternative time scales that were
derived in the previous development stages of the theory
(\citename{stenflo-s2018cc} \citeyear*{stenflo-s2018cc,stenflo-2020intech}) do not 
qualify as a dynamical time scale. In our earlier treatments it was not clear, 
whether or not there could be some coupling between 
the dark energy and dark matter problems. We now understand
that the two problems are not directly related. While this might seem
obvious in hindsight, it is
an important issue that needs to be clarified. In the following we
will outline the arguments, which made 
these conclusions unavoidable. 

For this it is sufficient to limit our focus on the BBN processes of
helium and deuterium formation, without the need to enter into
technical details. The arguments can be summarized as follows: If we
for instance enhance the expansion rate (by assuming that the
observer-induced edge effect is relevant for the non-equilibrium
physics), then $\Omega_B$ has to be raised to maintain
agreement with the observed deuterium abundance. In contrast,
$\Omega_B$ must be reduced to maintain agreement with the
observed helium abundance. Therefore it is impossible to satisfy both
the deuterium and helium constraints at the same time with a single
value for $\Omega_B$. 

The final deuterium abundance that is asymptotically approached when
all the other reaction rates have frozen out is primarily governed by the ratio
$R_{pD}/H$, where $R_{pD}$ is the rate for the $p+D \to\,  ^3{\rm
  He}+\gamma$ reaction \citeaffixed{stenflo-mukh2005}{cf.}, and $H$ is
the Hubble expansion rate. Because $R_{pD}\sim \Omega_B$, 
we need to keep the $\Omega_B/H$ ratio approximately unchanged to maintain
agreement with the observed deuterium abundance. This is the main
reason why an enhanced expansion rate requires a higher baryon fraction
according to the deuterium constraint. 

The final helium abundance on the other hand depends on the freeze-out
of the neutron abundance. This freeze-out depends on the expansion
rate $H$ but not on $\Omega_B$. An enhanced expansion rate leads to
earlier neutron freeze-out with higher neutron abundance. Normally all
these neutrons get used up to build helium. In this case enhanced
expansion rate would lead to excessive helium abundance. If however
the conversion to helium could be delayed by a time of order 10\,min
or longer, then only a fraction of the frozen-out neutrons would be
available for helium formation, because the rest would have
disappeared before due to spontaneous decay. As the rate of helium
formation scales with $\Omega_B$, the conversion process can be
slowed down by decreasing the value of $\Omega_B$. This
may be possible without violating the lower limit $\Omega_\star$ for the
allowed value of $\Omega_B$, where $\Omega_\star\approx 0.4\,$\%\ is
the estimated fraction of luminous matter (in stars) in the universe
\citeaffixed{stenflo-bookpeebles1993}{cf.}. As it is an order of magnitude smaller than the
nominal value for $\Omega_B$, there may be ample room for adjustments
of the value of $\Omega_B$ to slow down the rate of helium formation
as needed. 

While these examples show that it is possible to find independent fits
to the deuterium and helium abundances as long as the change in
expansion rate is not too large, a simultaneous fit to both
cannot be obtained. For large modifications of the expansion rate we enter
into different regimes, in which  even separate fits to the observed
helium or deuterium abundances cannot be found. The basic conclusion
is always the same: Any significant departure from the
expansion rate that is used in standard cosmology for the early 
(radiation-dominated) universe would lead to violation of the
joint observational constraints. As shown in the preceding section, the
expansion history that is described by the ``intrinsic'' Friedmann
model satisfies this requirement. It represents the
only viable version of the boundary-induced theory for the origin of
the cosmological constant, but it is a version that agrees with all the
constraints equally well as the concordance model. 

Note that the age of the universe according to the intrinsic Friedmann
model is 9.7\,Gyr, significantly less than the 13.8\,Gyr of the
concordance model. It is the shorter, ``intrinsic age'' that is the
physically relevant one. The age of the concordance
model is larger because it is affected by the presence of the
observer-induced $\Lambda$, which 
is an edge effect that is not relevant for the dynamical time scale. 

As a consequence of this analysis we can conclude that the dark matter
problem is not linked to the problem of the dark energy (cosmological
constant). The previously adopted low value for $\Omega_B$ is
unaffected. Most of dark matter needs to be non-baryonic to be
consistent with the observational data.

\section{Conclusions}\label{sec:concl}
In a world without participating observers there is no cosmological
constant, and the time line extends indefinitely into 
the future. This world is unobservable, because the introduction of an
observer changes its structure by causing a
split between past and future, concepts that do not
have any precise meaning in the observer-free world. Because the future is
fundamentally unobservable, the present represents a boundary, an edge
of time. The existence of this edge constrains the solutions. The
induced boundary effect is a repulsive force that surrounds the
observer with a region of accelerated expansion and a redshift
distribution that can be modeled with a cosmological constant. 

The boundary has no location in space, only in time (the ``Now''), as
if an infinitely high potential barrier has been erected to block
access to the future. The presence of this barrier disturbs the spacetime
metric, which in the cosmological context is represented by the scale
factor that scales with $1/(1+z)$, where $z$ is the redshift. The
disturbance is large for small redshifts, i.e., close to 
the edge, but vanishes with increasing distance from the boundary
(as the redshift increases). 

The value of the boundary-induced cosmological constant is independent
of redshift and has no gradients in the observable spacetime. It
therefore does not appear in the equation for energy conservation. As
an observer-induced effect it is not a physical field, but it affects
the way the universe appears from the perspective of the observer. The
relevant physics that governs the cosmic evolution, in particular 
the dynamical time scale,  is described by the Einstein equation for
an observer-free universe. It is represented by the ``intrinsic'' 
Friedmann model without a cosmological constant. The BBN and CMB
intrinsic time scales are indistinguishable 
from the time scale of standard cosmology, because the CMB
and BBN epochs are sufficiently far from the present temporal boundary to
be significantly affected by the cosmological constant in the
concordance model. The cosmology that follows from the
observer-induced interpretation of the cosmological constant therefore satisfies all
observational constraints equally well as the concordance model. In
particular this implies a reconfirmation of previous
conclusions that most of dark matter is non-baryonic. 

The formula Eq.~(\ref{eq:lamru}) for the cosmological constant,
$\Lambda =2\, (\pi/r_u)^2$, represents a resonance that is a consequence
of a periodic boundary condition for the finite region in Euclidean
(imaginary) time. In analogy with the wave-particle complementarity in
quantum physics, there is a complementarity between Euclidean
spacetime, with its wave-like world of complex phase factors, and
Lorentzian spacetime, with its particle-like world of evolving real
amplitudes. Both representations express complementary
aspects of time, which cannot be captured by
either representation alone. 

This leads us to a cosmological framework that does not contain any
``cosmic coincidence'' problem. There is nothing special about the
epoch in which we live, we are not privileged observers. There can be
no parallel universes with alternative values of the cosmological
constant, because logical consistency demands uniqueness.

\section*{References}



\end{document}